\documentstyle[11pt,newpasp,twoside,epsf]{article}
\def\edcomment#1{\iffalse\marginpar{\raggedright\sl#1\/}\else\relax\fi}
\reversemarginpar

\begin{document}
\title{Evidence for compact structuring in the corona of active stars}
 \author{F. Favata}
\affil{Astrophysics Division -- Space Science Department of ESA, ESTEC,
  Postbus 299, NL-2200 AG Noordwijk, The Netherlands}

\begin{abstract}

  The ``current wisdom'' regarding the structuring of the X-ray
  emitting corona in active stars (i.e.\ a corona dominated by
  extended coronal structures) is briefly reviewed, followed by a
  review of a new approach to flare analysis and the analysis of a
  significant number of newly observed and previously published large
  flares, all leading to a much more compactly structured corona.
  Recent observations showing the polar location of the flaring plasma
  are then discussed, showing how the current evidence points toward a
  (flaring) corona composed of rather low-lying polar structures, also
  in agreement with some recent radio VLBI observational results and
  with starspot Doppler images. The resulting picture is significantly
  different from the solar case.

\end{abstract}

\section{Introduction}

Active late-type stars have a coronal X-ray luminosity of up to $\ge
10\,000$ times higher than the Sun; it is thus natural to ask whether
the structuring of such a corona can be simply understood as a simple
extension of the solar case or not. Under structuring here we mean the
size, location and type of the coronal structures responsible for the
bulk of the observed emission measure at X-ray wavelengths (i.e.\ 
plasma at MK temperatures). For the Sun, high-resolution X-ray images
of the corona have allowed a detailed study of its spatial
structuring, showing a corona structured in magnetic loops confined to
intermediate latitudes, with ample coronal holes specially at the
solar poles. The typical maximum height of the corona above the
photosphere is well below the solar radius.  Can this corona be scaled
to one $10\,000$ times more luminous, and if so, how?


Stars are unresolved objects and therefore only indirect evidence can
be obtained about the spatial size, location and structuring of their
corona. Three basic types of observations can be used to derive
structural information about the corona, namely 1) the analysis of
flare decay, 2) the modulation of light due either to mutual eclipses
in a binary system or to self-eclipse due to stellar rotation in a
single star, and 3) the determination of plasma densities from
high-resolution spectra.

Here we briefly review the above approaches, showing that in many
cases the results have been interpreted, in the past, as giving
evidence for a large, rather diffuse corona. We then proceed to
discuss recent developments in the methodology as well as in the
observations, and discuss how these new development convincingly point
toward a picture of a rather compact corona, located on (or near) the
poles of active stars.

\section{Methodology}

\subsection{Flare decay analysis}

The analysis of the decay of flares can offer insights in the size and
characteristics of the flaring structures, and indeed it has often
been used in this way. Perhaps the most widely used approach for the
analysis of the decay of flaring loops has been the so-called
quasi-static approach (van\,den\,Oord \& Mewe 1989), which assumes
that the heated loops decay through a sequence of quasi-static states;
while the theoretical framework of the method includes sustained
heating as a free parameter, in practice its application has never
given any indication for the presence of heating during the decay
phase. As a consequence, when applied to the study of long-lasting
intense flares the approach always results in long, low-density loops,
which minimize both conductive and radiative losses, allowing the
plasma to cool slowly and thus justifying the observed long decay
times.

Another approach used at times has been the so-called two-ribbon
approach of Kopp \& Poletto (1984), which however requires a priori
assumptions on the spatial structure and thus is of limited
applicability in the stellar case.

More recently, an approach based on hydrodynamic modeling of the
decaying loops has been developed by Reale et al.\ (1997); this
approach uses the slope of the decay in the temperature-density plane
as an explicit diagnostic for the presence of sustained heating, and
has been tested on the Sun, showing that many apparently ``impulsive''
solar flares are actually dominated by sustained heating in the decay
phase. As a consequence, this approach results in the majority of
cases in smaller loops than the quasi-static approach.

One important point is that only intense (i.e.\ pathological in solar
terms) stellar flares yield sufficient statistics to allow a detailed
study, so that analogies with the solar case should be treated with
some caution, and deductions made on the Sun may or may not be a good
proxy for the stellar case. A significant number of intense events has
been observed by e.g.\ \emph{Einstein}, EXOSAT, ROSAT, ASCA and SAX.
Up to until ca.\ 1998 they have almost always been modeled with the
quasi-static approach, with the resulting model invariably pointing to
the presence of long loops. An example of some literature results
regarding the quasi-static analysis of large stellar flares is shown
in Table~1.

\begin{table}[tb]
\caption{Examples of some well known stellar flares. $L$ is the length
  of the structure obtained with the quasi-static approach.}
  \begin{tabular}{lll|lll}\tableline
   Star & Instrument & $L$ & Star & Instrument & $L$\\\tableline

Algol$^3$ & {\sc pspc} & { $\simeq 2.0~R_*$} & EV Lac$^1$ & {\sc pspc}
    & { $\simeq 10~R_*$} \\
    Algol$^2$ & {\sc exosat} & { $\simeq 0.6~R_*$} & YLW 15$^7$ & {\sc
    asca} & { $\simeq 3.0~R_*$}  \\ 
    Algol$^4$ & {\sc ginga} & { $\simeq 2.5~R_*$} & LkH$\alpha$ 92$^8$  &
    {\sc pspc} & { $\simeq 1.0~R_*$} \\
    AR Lac$^6$ & {\sc pspc} & { $\simeq 1.3~R_{\rm K}$} & V773 Tau$^9$
    & {\sc asca} & { $\simeq 1.2 R_*$} \\ 
    CF Tuc$^5$ & {\sc pspc} & { $\ge  2.7~R_{\rm K}$} & & & \\\tableline\tableline
  \end{tabular}
  \label{tab:lit}
  {\tiny $^1$Schmitt (1994), $^2$van den Oord et al.\ (1986), $^3$Ottman \&
  Schmitt (1996), $^4$Stern et al.\ (1992), $^5$K\"urster \& Schmitt (1996),
  $^6$Ottman \& Schmitt (1994), $^7$Tsuboi et al.\ (2000), $^8$Preibisch et
  al.\ (1993), $^9$Tsuboi et al.\ (1998).}
\end{table}


The purported very long loops populating the coronae of active stars
have led to a picture of an extended corona, which in some cases have
been pictured as extending between the components of binary systems
(e.g. the picture of Uchida \& Sakurai 1983), which has at times been
dubbed ``the standard model''.

\subsection{Eclipse mapping}

A spatially inhomogeneous corona will, when subject to the eclipse
from a companion, or to self-eclipse from the photosphere of the
parent star, produce a modulated light curve, which in principle could
be used to deduce its structure. The derivation of spatial information
from the observed light curve of either single or binary stars is a
process ripe with uncertainty, which makes a number of assumption
(e.g.\ that all observed variations are eclipse-induced, and not due
to the temporal variability of the emission) which are very fragile in
the presence of observational noise. The difficulties and significant
limitations of the approach have been discussed in detail by e.g.\ 
Schmitt (1998).  Nevertheless, a number of such analyses have appeared
in the past in the literature, notably on AR Lac (e.g.\ White et al.\ 
1990; Siarkowski et al.\ 1996); these have produced evidence for large
emitting regions, with the inter-binary region filled with plasma.
However, as the same authors also point out, other solutions are
equally possible (including a quite compact corona) if different a
priori constraints are assumed.

More in general, the observed lack of strong eclipses (e.g.\ on Algol)
has been interpreted as also pointing to the presence of very large
structures (loops), sufficiently larger than the parent star that only
a small fraction can be occulted at each given time, and thus in
agreement with the very large loops derived by flare decay analysis
through the quasi-static method. This view has contributed to a
general picture of active stellar coronae as quite extended objects.

The interpretation of light curves from active eclipsing binaries is
complicated by the difficulty of disentangling the contribution to the
light curve from each component. Binary systems with an X-ray active
star and a X-ray dark companions are indeed much better targets for
this type of studies, as the X-ray dark star acts as a pure occulting
disk. Examples of this type of stars are the Algol-type systems. A
pioneering study of such a system has been performed by Schmitt \&
K\"urster (1993): the clear observation of a sharp X-ray eclipse on
$\alpha$ CrB (a G5V$+$A0V system, $0.9 + 3.0~R_\odot$, $L_{\rm X}
\simeq 100\,L_{\rm X\odot}$) clearly indicates that the coronal
structures on the G5V star are compact, significantly smaller than the
star itself, with no evidence for extended structures.

\section{New observations of large stellar flares}

An observational breakthrough in the study of the structuring of
coronae in active stars was achieved in 1997 with a long observation
of Algol (a B8V$+$K3IV with a quiescent X-ray luminosity of $L_{\rm X}
\simeq 5000\,L_{\rm X\odot}$) performed with the SAX observatory. The
observation was dominated by a long-lasting (ca.\ 2 d), intense flare,
which underwent a total eclipse when the B-type star passed in front
of the K-type star.  This allowed Schmitt \& Favata (1999), for the
first time, to determine the size and location of an individual
coronal structure (the flaring ``loop'') in a deterministic way
(without the ambiguities present in active binaries). The results were
surprising, in that the flaring structure is rather compact (with a
maximum height $H<0.6\,R_*$) and located on the south pole of the
K-type star. Both the size and location of the flaring structure are
very different from the ``classic'' picture: there is no evidence for
extended structures nor for (near-) equatorial structures. In
addition, to allow for such a long-lasting event to take place in a
compact structure, the plasma must be heated throughout the decay
(else the natural decay time would be much faster than the observed
one).

\begin{figure}
\plottwo{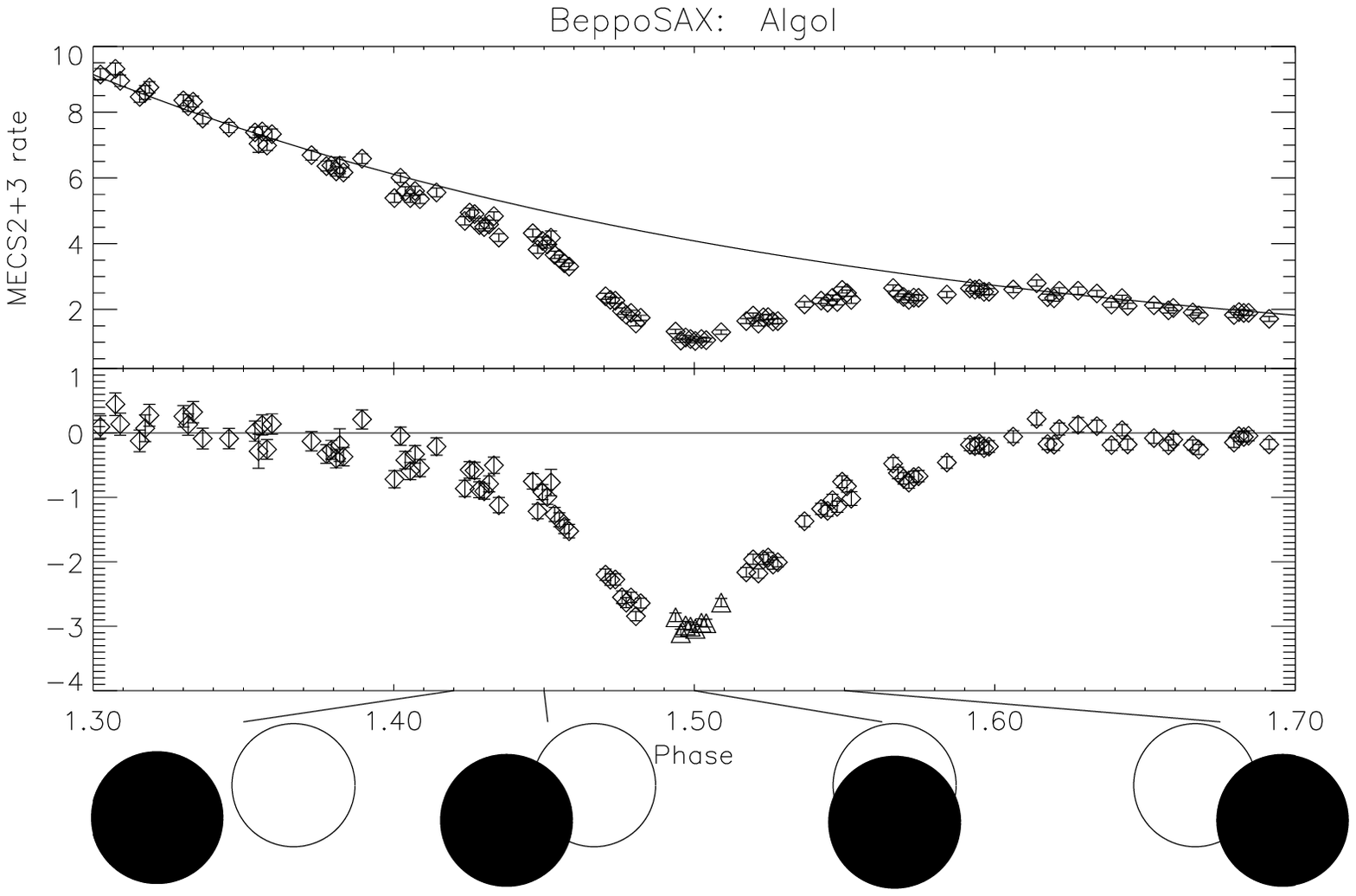}{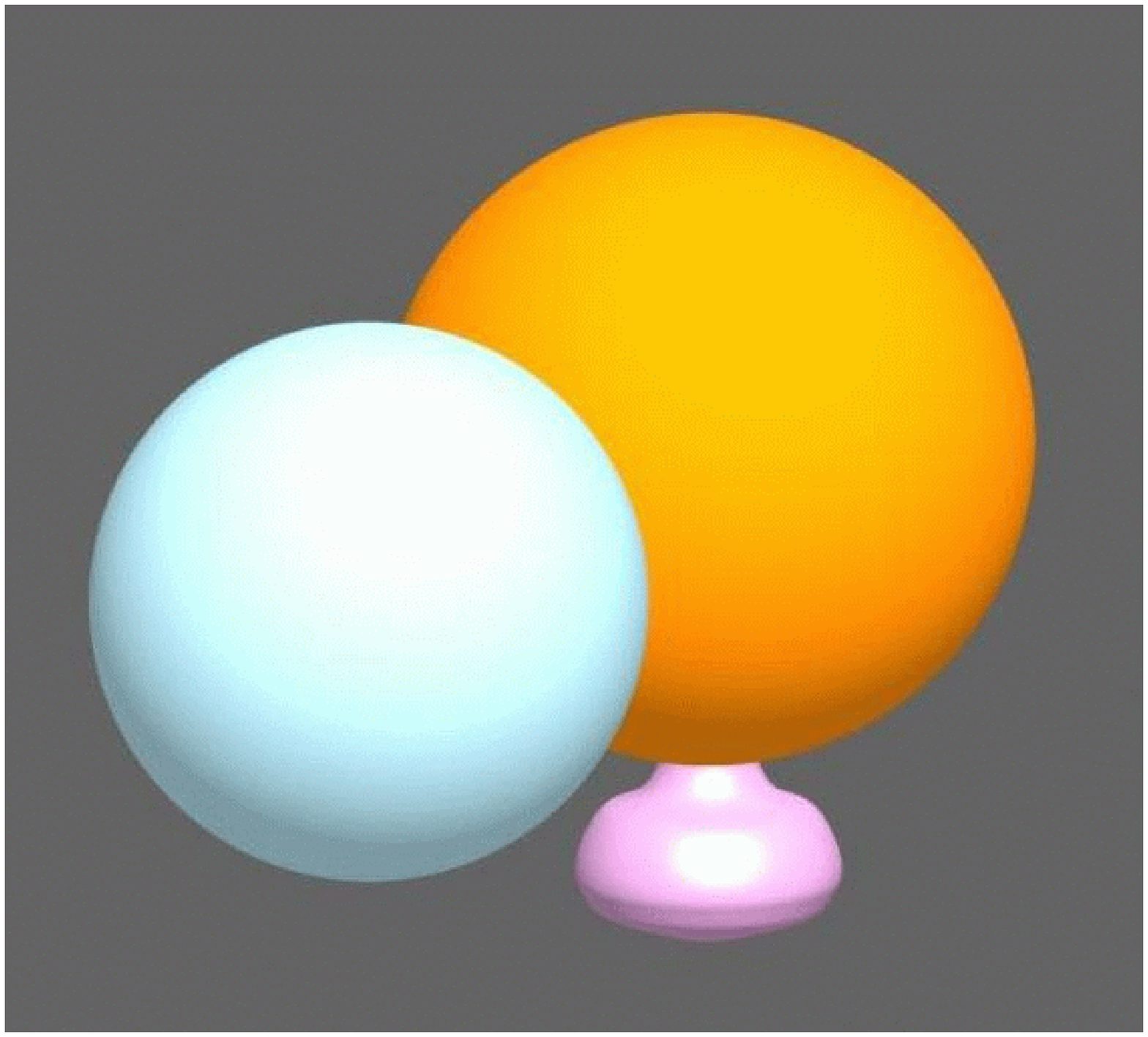}
\caption{The light curve of the SAX Algol flare eclipse is shown in
  the left-hand panel, together the relative position of the two
  stars. The right-hand panel shows the region (on the south pole of
  the K star) to which the flaring plasma must be confined to produce
  the observed eclipse light curve.}
\end{figure}

The excellent temporal coverage and statistics of this event have also
allowed Favata \& Schmitt (1999) to perform a detailed ``comparative
test'' of different flare-decay methods, for the first time having a
``ground thruth'' against which the reliability of the method could be
assessed.  The result is that the quasi-static method, when applied to
this event, overestimates the loop size by almost an order of
magnitude and fails to detect the presence of sustained heating. The
hydrodynamic method, on the other hand, properly diagnoses that the
loop must be heated during the decay, and thus predicts a smaller loop
size, closer to the ground truth (although still somewhat too large).

\subsection{A critical reassessment of the Algol corona}

Armed with the knowledge that the quasi-static method is likely to
consistently overestimate the size of the flaring structures, and that
the hydrodynamic method supplies a much better estimate, a systematic
re-analysis of all known large flares on Algol was performed by Favata
et al.\ (2000b). A summary of the results is shown in Table~2. The key
result of this work shows that in all cases strong sustained heating
is present, and that the flares are always relatively low-lying, with
no evidence for very long, extended loops.  This is in contrast with
the results originally obtained with the quasi-static method, which
claimed long loops and in all case freely decaying flares with no
evidence for sustained heating.

\begin{table}
  \caption{A summary of the reanalysis of the Algol flares performed
    by Favata et al.\ (2000b). The observed peak temperature
    ($T_{\rm max}$) and decay times are reported ($\tau_{\rm LC}$),
    followed by the ratio between the observed and intrinsic heating
    time ($\tau_{\rm LC}/\tau_{\rm th}$, indicating the importance of
    sustained heating), and by the length obtained with the
    quasi-static method ($L_{\rm QS}$) and with the hydrodynamic
    analysis ($L_{\rm Hy}$). For the SAX flare, the geometric size is also
    indicated.}
  \begin{tabular}{lrrrrrr}\tableline
    Instr. & $T_{\rm max}$ & $\tau_{\rm LC}$
    & $\tau_{\rm LC}/\tau_{\rm th}$& $L_{\rm QS}$ &
    \multicolumn{1}{c}{$L_{\rm Hy}$} & $L_{\rm geom}$\\
    & MK & ks & & $R_*$ & \multicolumn{1}{c}{$R_*$} & $R_*$ \\\tableline 
    EXOSAT     &  78 &  5.3 & 2.4 & 0.6 & {0.3} [0.2 0.4] & -- \\ 
    GINGA      &  67 & 19.8 & 4.2 & 2.4 & {0.5} [0.4 0.6] & -- \\  
    ROSAT      &  44 & 30.2 & 2.6 & 2.0 & {1.2} [1.1 1.3] & -- \\ 
    SAX        & 142 & 49.6 & 2.4 & 7.0 & 3.3 [1.9 4.7] &
    {0.9}\\\tableline \tableline  
    \end{tabular}
\label{tab:algol}
\end{table}

\subsection{Other examples: flare stars}

A similar type of analysis, with a consistent analysis of all known
flaring events, was performed by Favata et.\ al (2000a) on the
``prototypical'' flare star AD Leo. The significant number of events
studied, across different instruments and detectors, allows to deduce
some general conclusions: the characteristic size of the flaring
structures is small, well below the stellar radius, and vigorous
sustained heating is in essentially all cases present. The
characteristic loop size is, for AD Leo, even smaller than for Algol,
with a characteristic size of $L \simeq 0.3 R_*$.

Also, an exceptional flare (whose total energy was, at peak, as large
as the stellar photospheric luminosity) was detected with ASCA on the
flare star EV Lac (see Fig.~2). Notwithstanding the exceptional
intensity of the event, the hydrodynamic analysis (Favata et al.\ 
2000d) of the event shows that, similarly to the Algol SAX flare,
the flaring plasma is confined to a region $L \le 0.5\,R_*$, and that
the decay light curve is also dominated by the time evolution of the
sustained heating. 


\subsection{Other examples: active binaries}

Similar conclusions have been reached (Favata et al.\ 2001) on active
binaries through the analysis of some well-known events previously
studied in the literature. For example, for the flare observed by
ROSAT on AR Lac (Ottmann \& Schmitt 1994) a quasi-static analysis
yields a size larger than the K star in the system; this is reduced to
$L \le 0.5\,R_*$ when analyzed with the hydrodynamic method.
Similarly, the long and intense event observed by ROSAT on CF Tuc
(K\"urster \& Schmitt 1996) for which the quasi-static analysis yields
a size for the flaring region of $L \simeq 3\,R_*$ is reduced to $L
\le R_*$ by the hydrodynamic analysis. Once more, the decay of these
large flaring events is dominated by the presence of vigorous
sustained heating.

\subsection{Other examples: pre-main sequence stars}

A re-analysis of large flaring events on pre-main sequence (PMS) stars
(Favata et al.\ 2000c) also shows that sustained heating during the
decay phase is a common feature and that the size of the flaring
regions obtained with the hydrodynamic analysis is significantly
smaller than the one obtained previously with the quasi-static
approach. In general, the ``larger than the star'' loops found by the
quasi-static analysis are reduced to ``smaller than the star''
structures. This applies to different categories of PMS stars; for the
ASCA flares on the proto-star (YSO) YLW 15 the analysis of Tsuboi et
al.\ (2000) found a large loop ($L \simeq 3\,R_*$), which is reduced
to $L \le R_*$ by the hydrostatic analysis. Similarly, for the flare
observed on the classical T Tau star (CTTS) LkH$\alpha$~92 by
Preibisch et al.\ (1993) the size shrinks from $1$ to $0.5\,R_*$,
while for the ASCA flare on the weak-line T Tau (WTTS) star V773 Tau
(Tsuboi et al.\ 1998) the size shrinks from $1.2$ to $0.3\,R_*$.
Finally, an analysis with the hydrodynamic method of the ROSAT flare
on the WTTS HD~283572 reported by Stelzer et al.\ (2000) yields a size
for the flaring structure of $\simeq 0.4\,R_*$.  Also, an analysis
with the same approach of the ``twin'' large flares observed by SAX on
the zero-age main sequence (ZAMS) star AB Dor (Maggio et al.\ 2000)
consistently results in structures well below the stellar size.  In
general, the consistent small size found for the flaring structures on
PMS stars casts a doubt on the reality of the magnetic structures
extending from the star to the accretion disk which have been invoked
(e.g.\ Montmerle et al.\ 2000) to explain the events on YSO's and
CTTS's. The general characteristics of these events are
indistinguishable from the ones taking place on single main sequence
stars (e.g.\ on AB Dor), thus pointing toward similar coronal
structures being present in both cases.

\section{Discussion}


\subsection{Size of flaring structures}

The wealth of new observations (and re-analysis of existing
observations) discussed above shows consistently that application of
the hydrodynamic method (whose superiority was clearly shown on the
Algol SAX flare) results in much smaller structures than the ones
obtained with the quasi-static analysis (whose failure at detecting
sustained heating was, again, shown on the Algol SAX flare) of the
same events. In general, all flaring structures are smaller than the
size of the star, with sustained heating being a common feature.  The
size of the flaring coronal structures derived with hydrodynamic
modeling are in general, by solar standards, large, but not
``exceptional''. The presence of strong sustained heating makes it
likely that even the hydrodynamic method actually over-estimates the
size of the coronal structures, so that, if anything, the actual size
is likely to be even smaller. These conclusions apply to active stars
spanning a wide range of mass, sizes and evolutionary stages.

\subsection{Time-profile of the sustained heating}

One consequence of the presence of sustained heating during the decay
of flares is that the observed light curve is dominated by the
time-profile of the heating process and does not reflect the actual
structure of the flaring structure. Thus, (repeated) features in the
decay light curve should provide hints to the flare heating mechanism.
One feature which appears to be consistently present in intense flares
(when the temporal resolution and the $S/N$ are sufficient) is the
presence of a ``knee'' in the decay light curve: the initial
(exponential) decay is fast, and it slows down significantly in the
second part of the event. This is apparent, e.g.\ in the two events
shown in Fig.~2. Similar features are visible in intense solar events.

\begin{figure}
\plottwo{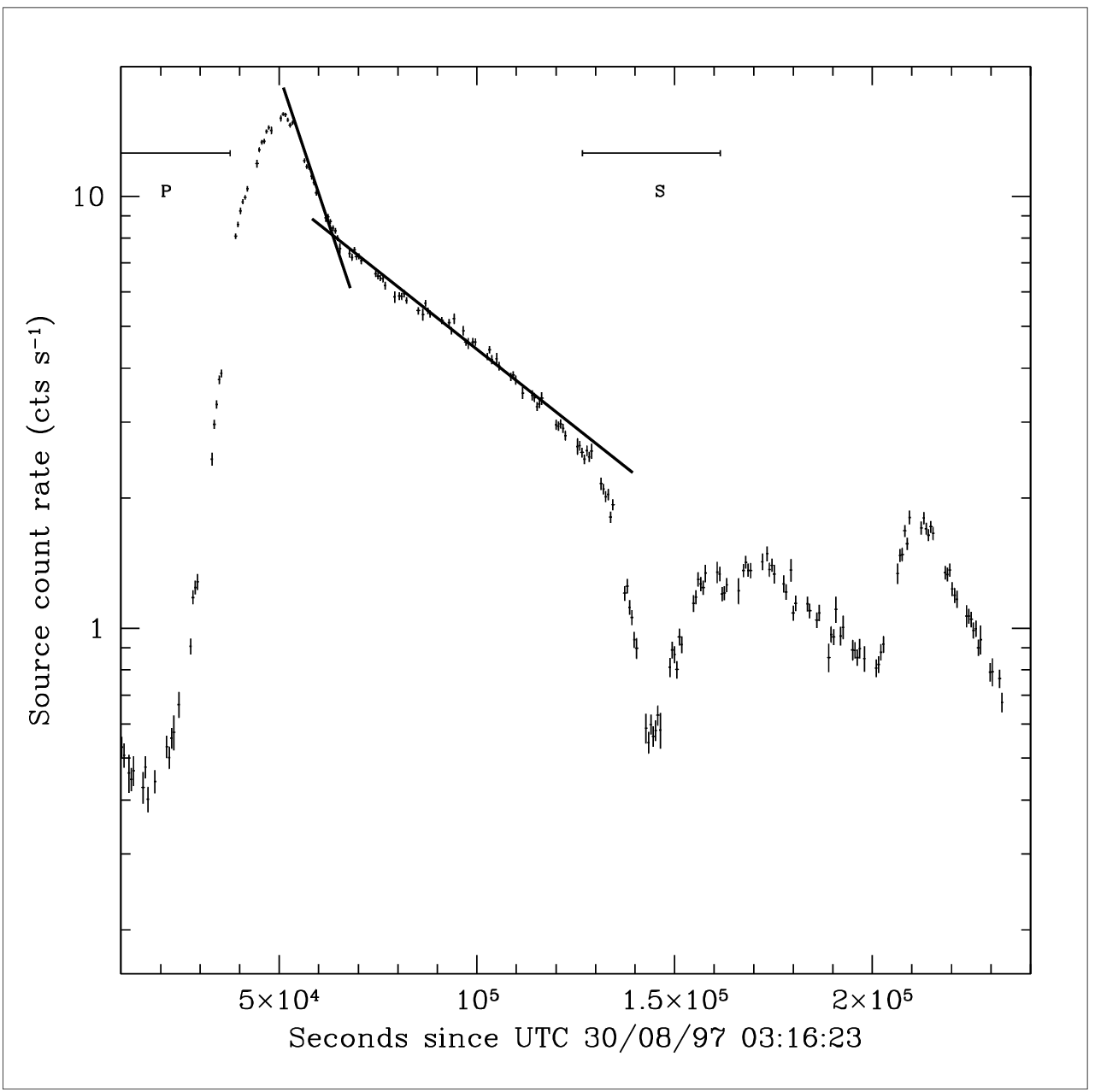}{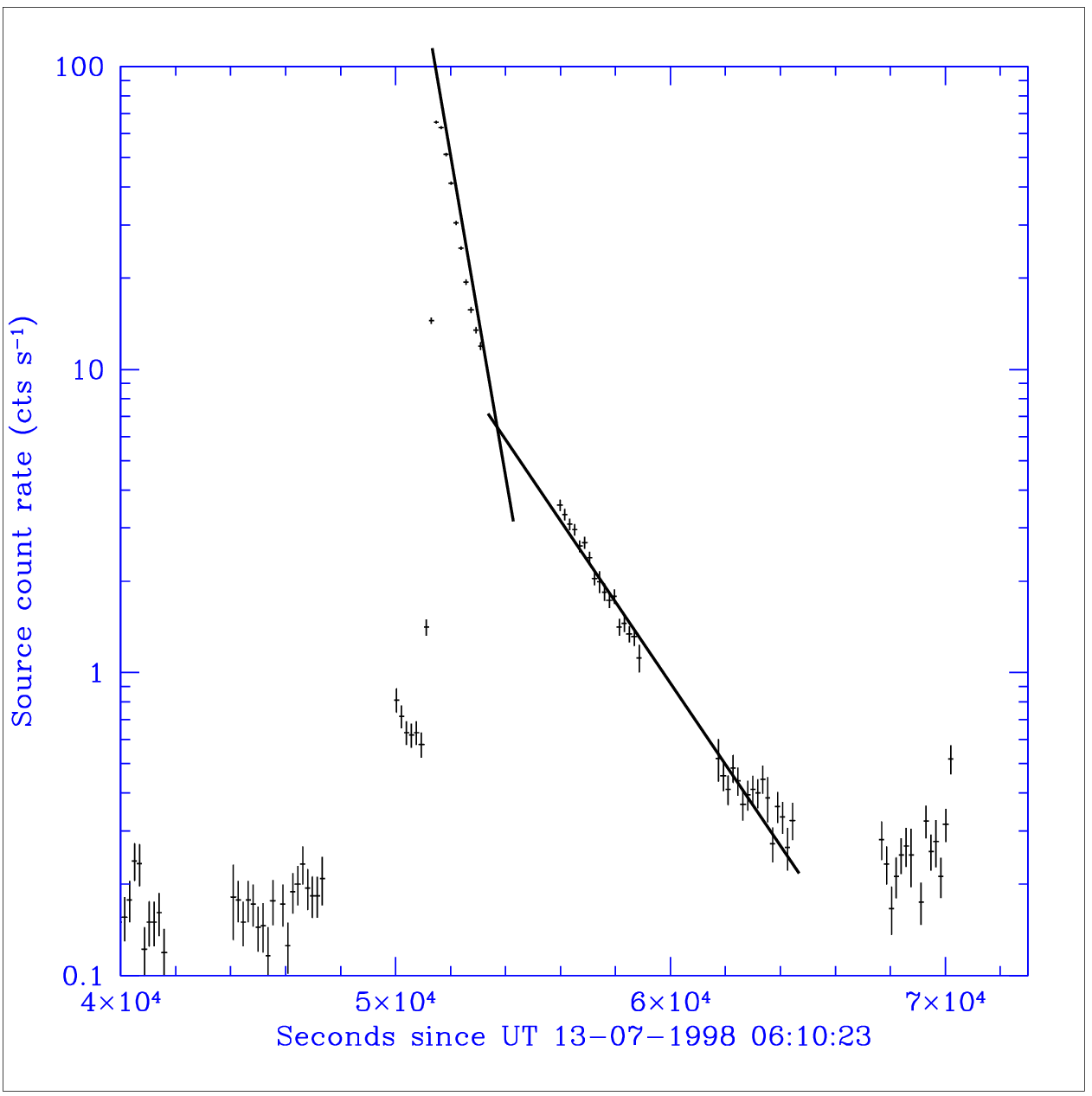}
\caption{The decay light curve of the SAX Algol flare (left-hand panel) and
  of the ASCA EV Lac flare (right-hand panel) showing in both cases an
  initially steep decay followed by a slower phase.}
\end{figure}

\subsection{Location of the flaring plasma}

The (circumstantial) evidence about the location of the flaring plasma
available prior to the Algol SAX flare has in general been interpreted
as pointing toward structures located at low stellar latitudes. The
two key lines of evidence used for this argument have been the results
of the light-curve deconvolution and the long duration of large
flaring events. 


Light curve deconvolution of active binaries (e.g.\ Siarkowski et al.
1996) results mostly in coronal structures located at low latitudes
(near-equatorial structures); however, as discussed for the size of
the coronal structure, the results are quite sensitive to the a priori
constraints assumed, so that it is unclear whether solutions with
(predominantly) high-latitude structures would be equally possible.

Intense stellar flares often have a duration comparable to (or even
longer than) the stellar rotational period (or the orbital period in
the case of binary systems). The lack of observed self (or mutual)
eclipses of these long events, coupled with the inference (from
quasi-static analyses) that the flaring structures where very large,
had led to the deduction that they where likely located close to the
system's equator. The large size was used to justify the observed lack
of eclipses. The observed eclipse of the Algol SAX flare, together
with the evidence for predominantly compact flaring structures
discussed above, shows that these deductions are unlikely to be
correct. The Algol flare results from a polar structure, and a polar
location is also compatible (and indeed necessary) to explain many
other flare observations. The flares observed with SAX on AB Dor by
Maggio et al.\ (2000) last for a time comparable to the stellar
rotational period, and yet the flaring structures are small and are
not self-eclipsed. This can only be explained if the flaring
structures are located on the (exposed) stellar pole, so that they are
never eclipsed, even if small in size. A similar argument can be used
to explain other similar events: the very long-lasting flare observed
by ROSAT on the flare star EV Lac (Schmitt 1994) lasts for more than
one stellar rotation, yet it is not self-eclipsed. Given that even
much more intense flares (e.g.\ the ASCA event studied by Favata et
al.\ 2000d) on the same star can be easily explained as confined in
small structures, once more only a polar location can explain the
observed lack of self-eclipses. Similarly, the ASCA flares observed on
the YSO YLW 15 last for a time comparable with a rotational period,
yet they appear to be confined in small regions and not self-eclipsed.
Again, a polar location appears to be the only one possible. One
observation of YLW 15 (Tsuboi et al.\ 2000) three flares have been
seen with a recurrence time comparable to the stellar rotation period.
This has been interpreted as due to rotational modulation of the
magnetic field stress; no rotational modulation of the light curve (as
due to self-eclipse) is however visible. Such interpretation is thus
not per se in contradiction with a polar location of the flaring
plasma.

A (predominantly) polar location can also naturally explain the modest
amounts of modulation observed in active binaries, and is compatible
with the predominantly photospheric polar large spots deduced by
Doppler imaging. Some theoretical models indeed predict that, for
rapid rotators, the poles are the preferential location for the
emergence of magnetic flux.

Interestingly, recent radio VLBI observations of Algol by Mutel et al.
(1998) show that the (radio) flaring corona is also coming from two
polar lobes with characteristic sizes smaller than the size of the K
star in Algol, while the quiescent emission comes from a somewhat more
extended region, comparable in size to the star, although also located
on the polar regions. As shown by Favata et al.\ (2000b), a similar
structure for the X-ray active corona can also explain the known
characteristics of the X-ray emission. However, the polarization of
the radio emission shows that the large scale field is likely bipolar,
so that the solar-like ``loop'' picture may be inappropriate for the
Algol corona (as well as for other high-activity ones), and that
different type of magnetic structures may be necessary to confine the
plasma.

\section{Conclusions}

The large body of observational evidence presented here (and
accumulated mostly in the last two years) shows that the active
(flaring) component of the corona of active stars is confined in
structures which are significantly smaller than the star itself, and
which are likely located at high stellar latitude. This is in contrast
with the picture of very long, near-equatorial coronal structures
(even with interconnecting loops in the case of active binaries, e.g.\ 
Uchida \& Sakurai 1983) which has in the past been proposed to
describe active coronae. Independent evidence for material at
significant distance from the stars, and in the inter-binary region
comes from Doppler mapping done e.g.\ with IUE observations (Pagano et
al.\ 2000, on AR Lac), using chromospheric lines. Such evidence is not
in contrast with the picture presented here, given than the two
concern material at very different temperatures, and thus likely not
co-located: the flaring coronal plasma discussed here is mostly at
temperatures of several MK to several tens of MK, while the
chromospheric material whose presence is deduced through Doppler
analysis (using e.g.\ Mg\,{\sc ii} lines) is at temperatures $T \le
10^5 K$. Similar considerations apply to the evidence for large
structures orbiting the young star AB Dor obtained by Collier Cameron
et al.\ (1990): also in this case, the material detected is at much
cooler temperatures than the coronal material responsible for the
X-ray (flaring) emission, and thus unlikely to be spatially coincident
with it.

The polar location of the flaring plasma is in stark contrast with the
solar case, where active regions are confined to low and intermediate
latitudes. This (together with the evidence from the radio
observations of a largely dipolar field) points to a magnetic field
structure, in the case of active stars, significantly different from
the solar one, and thus also possibly to a different
dynamo. Therefore, for the very active stars, the solar analogy may
have to be considered with some caution, and the coronal structures
responsible for the bulk of the (flaring) X-ray emission may indeed
look different than in the Sun.

The paradigm of a compact (and polarly located) corona in active stars
is also supported by the EUVE observations of contact (W UMa-type)
binaries, for which high density are deduced from spectroscopy (and
thus small volumes) for the emitting plasma, and the lack of strong
orbital modulation (as well as the period difference with respect to
the orbital motion, Brickhouse \& Dupree 1998 on 44 Boo) also point to
a polar compact corona. The upcoming flow of grating spectral
observations of active stars from Chandra and XMM-Newton will likely
put to test the picture presented here: spectroscopic observations of
the decay of an intense flare, once achieved, would allow to
independently measure the density and compare it with the one deduced
from the hydrodynamic method used here. The ``holy grail'' for the
field of coronal structuring would however be a grating observation of
an event similar to the SAX Algol one.



\begin{references}

  Brickhouse, N. S. \& Dupree, A. K. 1998, \apj, 502, 918

  Collier Cameron, A., Duncan, D. K., Ehrefreund, P. et al.\ 1990,
  \mnras, 247, 415

  Favata, F. \& Schmitt, J. H. M. M. 1999, \aap, 350, 900

  Favata, F., Micela, G., Reale, F. 2000a, \aap, 354, 1021

  Favata, F., Micela, G., Reale, F., Sciortino, S., Schmitt, J. H. M.
  M. 2000b, \aap, in press

  Favata, F., Micela, G., Reale, F. 2000c, submitted

  Favata, F., Reale, F., Micela, G. et al.\ 2000d, \aap, 353, 987

  Favata, F., Micela, G., Reale, F. 2001 in preparation

  Kopp, R. A. \& Poletto, G. 1984, Sol. Phys., 93, 351

  K\"urster, M. \& Schmitt, J. H. M. M. 1996, \aap, 311, 211

  Maggio, A., Pallavicini, R., Reale, F., Tagliaferri, G. 2000, \aap,
  356, 627

  Montmerle, T., Grosso, N., Tsuboi, Y., Koyama, K. 2000, \apj, 529,
  1097

  Mutel, R. L., Molnar, L. A., Waltman, E. B., Ghigo, F. D. 1998,
  \apj, 507, 371

  van den Oord, G. H. J. \& Mewe, R. 1989, \aap, 213, 245

  Ottmann, R., Schmitt, J. H. M. M. 1994, \aap, 283, 871

  Pagano, I. et al., 2000, \aap, in press

  Preibisch, Th., Zinnecker, H., Schmitt, J. H. M. M. 1993, \aap, 279,
  L33

  Reale, F., Betta, R., Peres, G., Serio, S., McTiernan, J. 1997,
  \aap, 325, 782

  Schmitt, J. H. M. M. 1994, \apjs, 90, 735

  Schmitt, J. H. M. M. 1998, in ``Cool Stars, Stellar Systems and the
  Sun'', ASP 154, 463

  Schmitt, J. H. M. M. \& Favata, F. 1999, Nature, 401, 44

  Schmitt, J. H. M. M. \& K{\"u}rster, M. 1993, Science, 262, 215

  Siarkowski, M., Pres, P., Drake, S. A., White, N. E., Singh, K. P.
  1996, \apj, 473, 470

  Stelzer, B., Neuh{\"a}user, R., Hambaryan, V. 2000, \aap, 356, 949

  Tsuboi, Y., Koyama, K., Murakami, H. et al.\ 1998, \apj, 503, 894

  Tsuboi, Y., Imanishi, K., Koyama, K., Grosso, N., Montmerle, T.
  2000, \apj, 532, 1089

  Uchida, Y. \& Sakurai, T. 1983, in ``IAU Colloq. 71: Activity in
  Red-Dwarf Stars'', 629

  White, N. E., Shafer, R. A., Horne, K., Parmar, A. N., Culhane, J.
  L. 1990, \apj, 350, 776

\end{references}

\end{document}